\documentclass[aps,prd,preprintnumbers,superscriptaddress,nofootinbib,amsmath,amssymb,showpacs]{revtex4}
\usepackage{dcolumn}
\usepackage{bm}
\usepackage{amsmath}
\usepackage{color}

\usepackage{enumerate}
\usepackage{epsfig}
\usepackage{overpic}
\usepackage{graphicx,floatflt,subfigure}

\usepackage{subfigure}

\usepackage{boxedminipage}

\input{colordvi.tex}

\newcommand{\beq}{\begin{equation}}
\newcommand{\eeq}{\end{equation}}
\newcommand{\beqa}{\begin{eqnarray}}
\newcommand{\eeqa}{\end{eqnarray}}

\newcommand{\cala}{{\cal A}}
\begin{document}

\title{Topological inflation from the Starobinsky model in supergravity}

\author{Kohei Kamada}
\email[Email: ]{kohei.kamada"at"epfl.ch}
\affiliation{
\it Institut de Th\'eorie des Ph\'enom\`enes Physiques,
\'Ecole Polytechnique F\'ed\'erale de Lausanne,
CH-1015 Lausanne, Switzerland}

\author{Jun'ichi Yokoyama}
\email[Email: ]{yokoyama"at"resceu.s.u-tokyo.ac.jp}
\affiliation{Research Center for the Early Universe (RESCEU), Graduate School of Science,
The University of Tokyo, Tokyo 113-0033, Japan}
\affiliation{Kavli Institute for the Physics and Mathematics of the
Universe 
(Kavli IPMU), TODIAS, WPI, 
The University of Tokyo, Kashiwa, Chiba, 277-8568, Japan}

\preprint{ RESCEU-11/14}

\begin{abstract}
We consider the ghost-free higher-order corrections to the Starobinsky model 
in the old-minimal supergravity, 
focusing on a sector among several scalar fields in the model 
that reproduces the scalaron potential in the original Starobinsky model. 
In general, higher-order corrections cannot be forbidden by symmetries, 
which likely violate the flatness of the scalaron potential 
and make inflation difficult in explaining the present Universe. 
We find a severe constraint on the dimensionless coupling of the $R^4$ correction as 
$-5.5 \times 10^{-8}<s<9.1 \times 10^{-8}$ from the recent results of the Planck observation. 
If we start from the chaotic initial condition, the constraint becomes much more severe. 
However, in the case in which 
the coupling of the $R^4$ correction is positive, the scalaron potential 
has a local maximum with 
two local minima at the origin and infinity, which admits topological inflation. 
In this case, inflation can take place naturally if the coupling satisfies the observational constraints. 
\end{abstract}

\maketitle

\section{Introduction}

The curvature-square inflation originally proposed by Starobinsky 
\cite{Starobinsky:1980te} occupies a unique position in inflationary
cosmology \cite{Guth:1980zm} because it only requires a single
additional term in the Einstein--Hilbert action instead of a new
scalar field---a functional degree of freedom.  Despite its simplicity, 
$R^2$ inflation is fully consistent with the state-of-the-art 
cosmological observations of the cosmic microwave background (CMB)
by WMAP \cite{Hinshaw:2012aka}
and Planck \cite{Ade:2013uln}, which report the scalar spectral index 
$n_s=0.963 \pm 0.014$ \cite{Ade:2013uln}, 
tensor-to-scalar ratio   $r\lesssim 0.135$  \cite{Ade:2013uln,Audren:2014cea},  
and the $f_{\rm NL}$ parameter measuring possible deviation from Gaussianity of curvature 
perturbations being consistent with zero.

Recently, the BICEP2 experiment \cite{Ade:2014xna}
announced the detection of B-mode polarization of the 
CMB on relatively low multipoles with its amplitude corresponding to 
$r \sim 0.2$.  This is an epoch-making discovery if it is confirmed to 
be due to the primordial gravitational waves, but at the moment, the
possibility that the detected signal is entirely due to polarized dust
has been ruled out only at 2.2$\sigma$ level \cite{Ade:2014xna}. Furthermore, it has been
pointed out recently that the effect of foreground dust may have been
even larger, so that only an upper bound on $r$ can be obtained as reported in Ref.~\cite{seljak}.  
In this sense, we had better not conclude in haste that
 $R^2$ inflation, which predicts
$r \simeq 3\times 10^{-3}$, has been ruled out by the latest observations of
B-mode polarization.

Turning our attention to more theoretical aspects, this model is so simple
that it had not been investigated in the context of 
modern high-energy theory including supersymmetry and supergravity, 
which would be the only theories that allow us to use the usual perturbative quantum field theory 
without a fine-tuning up to
scales relevant to inflation and necessary to embed it in the most promising candidate of the quantum
theory of gravity, the superstring theory.  
It is only recently that $R^2$-type inflation was studied based on 
supergravity 
\cite{Kallosh:2013lkr,Farakos:2013cqa,Ferrara:2013kca,Hamaguchi:2014mza}\footnote{Another model of supergravity 
extension of the Starobinsky model known as $F(R)$ supergravity 
has also been proposed  \cite{Ketov:2009wc} and its cosmological consequences have been studied 
\cite{Watanabe:2013lwa}. 
But its insufficiency as a supersymmetric theory has been pointed out recently \cite{Ferrara:2013wka,Ketov:2013dfa}. 
See also  \cite{Ellis:2013xoa} for
supersymmetric models that have a potential similar to the Starobinsky model.}, 
although its basic framework was already known 
in the late 1980s using the old-minimal supergravity \cite{Cecotti:1987sa} or the new-minimal supergravity 
\cite{Cecotti:1987qe}, which were obtained by the gauge fixing of the superconformal theory.

One of the most important messages of these studies is that higher-order corrections such as 
the $R^4$ term will arise as the nonrenormalizable operators. 
The effect of the $R^4$ term on the Starobinsky model can be seen easily when we go to the scalaron picture, 
which is the dual theory of $f(R)$ theory that consists of a scalar field and the Einstein--Hilbert action. 
In the scalaron picture, the Starobinsky model has a very flat potential at larger field values that is 
suitable for inflation, but $R^4$ correction destroys the flatness of the potential. 
Therefore, it must be strongly suppressed for the successful inflation \cite{Farakos:2013cqa,Ferrara:2013kca}. 

In this paper, we investigate the quantitative constraint on the $R^4$ corrections to the Starobinsky model
in supergravity in light of the observational result and the possibility of its realization. 
Even if the scalaron potential admits an inflationary solution, it is nontrivial whether it generates 
the primordial perturbations consistent with the current observation. 
As a result, we find severe constraints on the amplitude of the $R^4$ term. 
Moreover, it is also nontrivial how severe tuning for the initial condition is required, 
since the regions for successful inflation in the field space are drastically limited. 
We find that the coupling of the $R^4$ term  is very severely constrained if we start from the chaotic initial condition 
\cite{Linde:1983gd}. However, we also find that if the scalaron potential vanishes at the larger field values, 
depending on the sign of the coupling of the $R^4$ corrections, topological inflation \cite{Linde:1994hy,Sakai:1995nh,Ellis:1998gf} would be possible. 
Therefore, the initial condition problem is solved in this case, and we should only focus on the observational constraint 
for the embedding of the Starobinsky model in a supersymmetric theory. 
Here, we take the old-minimal supergravity for concreteness, but 
the same result is obtained in other supersymmetric extensions such as 
the new-minimal supergravity, too, as far as 
the form of the scalaron potential is concerned after all the other degrees of freedom have been stabilized. 
Note that the observational constraint for the $R^4$ correction to the nonsupersymmetric Starobinsky model
has been studied in Ref.~\cite{Huang:2013hsb}. Our result is slightly different but  consistent. 

The paper is organized as follows. 
In Sec.~\ref{sec2}, we derive the scalaron potential of the Starobinsky model in old-minimal supergravity 
with $R^4$ correction.  In Sec.~\ref{sec3}, we show the observational constraints to the $R^4$  correction 
from the Planck results. We examine the initial condition problem from the chaotic initial condition and 
show that the topological inflation likely takes place in Sec.~\ref{sec4}. Section~\ref{sec5} is devoted to the summary. 
In the Appendix \ref{app}, we show the equivalence of the scalaron potential to the nonsupersymmetric Starobinsky model. 

\section{Starobinsky model in old-minimal supergravity with higher-order corrections \label{sec2}}

We start from the old-minimal supergravity from superconformal gravity  \cite{Cecotti:1987sa}, 
where the conformal symmetry is broken by the gauge fixing of a chiral compensator field\footnote{
In the new-minimal supergravity \cite{Cecotti:1987qe}, the conformal symmetry is broken by a real linear multiplet.}
following the 
discussion of Ref.~\cite{Farakos:2013cqa}. 
Here we introduce a chiral compensator superfield, $S_0$, for which the scaling weight is 1 and 
chiral weight is 1/2. 
Action in the superconformal theories can be categorized by the D-type and F-type Lagrangians. 
The D-term Lagrangian is expressed as
\begin{equation} 
{\cal L}_D=[V]_D=\int  d^2 \Theta {\cal E}P[V]+{\rm h.c.}, 
\end{equation}
where $V$ is a real function of chiral fields for which the 
scaling weight is 2 and chiral weight is 0, 
${\cal E}$ is the chiral measure, and 
$P$ is the chiral projector in conformal superspace. 
The F-term Lagrangian is expressed as 
\begin{equation}
{\cal L}_F+{\rm h.c.}=[W]_F+{\rm h.c.}=\int  d^2 \Theta 2{\cal E} W +{\rm h.c.}, 
\end{equation}
where $W$ is a holomorphic function for which the scaling weight is 3 and chiral weight is 2. 

The gravity part of the standard supergravity Lagrangian is obtained by the compensator chiral superfield as
\begin{equation}
{\cal L}=-3 [S_0{\bar S}_0]_D
\end{equation}
with the gauge fixing $S_0=1$\footnote{Here, we take the reduced Planck mass $M_{\rm pl}=1$.}. 
Introducing a chiral multiplet
\begin{equation}
{\cal R}\equiv \frac{1}{2} S_0^{-1}  P[{\bar S}_0], 
\end{equation}
for which the scaling weight is 1 and chiral weight is 2/3, we obtain the $R^2$ correction to the standard supergravity Lagrangian, 
\begin{equation}
{\cal L}=-3[S_0{\bar S}_0]_D+3 \lambda_1[{\cal R} {\bar {\cal R}}]_D. \label{orig}
\end{equation}
Note that after gauge fixing, the chiral projector is expressed as 
\begin{equation}
P[V]=-\frac{1}{4}({\bar {\cal D}}{\bar {\cal D}}-8{\cal R}) V, 
\end{equation}
where ${\cal D}$ is the covariant derivative,  
and the ${\cal R}$ is the curvature chiral superfield. 
As a result, the Lagrangian reads 
\begin{equation}
{\cal L}=-3 \int d^2 \theta 2 {\cal E}  \left\{{\cal R}+\frac{\lambda_1}{8}({\bar {\cal D}}{\bar {\cal D}}-8{\cal R})({\cal R} {\bar {\cal R}})\right\} +{\rm h.c.},
\end{equation}
which has been shown to be the supersymmetrization of the Starobinsky's $R^2$ model \cite{Ketov:2013dfa,Cecotti:1987sa}. 

Now, let us consider higher-order corrections without fixing $S_0=1$ for the moment. 
Since $[f({\cal R} {\bar {\cal R}})]_D$ corrections have been found not to lead the supersymmetrization of 
$R^n$ corrections \cite{Ketov:2013dfa}, 
here we focus on the ghost-free correction that includes covariant derivatives of the curvature multiplet \cite{Khoury:2010gb},  
\begin{equation}
\Delta {\cal L}=\xi \left[ \nabla^\alpha ({\cal R}/S_0)  \nabla_\alpha ({\cal R}/S_0){\bar \nabla}_{\dot \alpha} ({\cal {\bar R}}/{\bar S}_0)  {\bar \nabla}^{\dot \alpha} ({\cal {\bar R}}/{\bar S}_0)  \right]_D, \label{corrrrr}
\end{equation}
where $\nabla$ represents the covariant derivative in the superconformal theory.  
As we will see, the Lagrangian contains only first derivatives of the fields 
and all the terms with the form of ${\cal L} \propto (\partial_\mu \phi)^2$ have the correct signs along 
the inflationary trajectory. Thus,
this system does not suffer from the emergence of the ghost degrees of freedom. On the contrary,
terms expressed with a function $g$ as $[g(S_0^{-1}{\cal R}, {\bar S}_0^{-1}{\bar{\cal R}}, S_0^{-2}P({\bar {\cal R}}), {\bar S}_0^{-2} {\bar P}({\cal R})) S_0 {\bar S}_0]_D$ give ghost degrees of freedom \cite{Cecotti:1987sa,Ketov:2013dfa}, and hence 
we do not consider them here. 
Therefore, Eq.~\eqref{corrrrr} would be the lowest nonrenormalizable term that can give the consistent theory 
in this framework. 

Now, we examine the structure of this system by using the Lagrange multiplier method, or the scalaron picture. 
By introducing a chiral superfield $\cala$ with scaling weight 1 and chiral weight 2/3, and a chiral Lagrange 
multiplier superfield $\Lambda$ with scaling weight 2 and chiral weight 4/3, the Lagrangian can be rewritten as
\begin{align}
{\cal L}&=-3[S_0{\bar S}_0]_D+3 \lambda_1[\cala{\bar {\cala}}]_D+\xi \left[ \nabla^\alpha (\cala/S_0)  \nabla_\alpha ({\cala}/S_0)  {\bar \nabla}_{\dot \alpha} ({\bar \cala}/{\bar S}_0) {\bar \nabla}^{\dot \alpha} ( {\bar \cala}/{\bar S}_0) \right]_D+3[\Lambda (\cala-{\cal R})]_F+{\rm h.c.}
\end{align}
By integrating out the multiplier field, we have $\cala={\cal R}$ and the original Lagrangian Eqs.~\eqref{orig} and \eqref{corrrrr}
are reproduced. 
Noting that the identity $[\Lambda {\cal R}]_F+{\rm h.c.}=(1/2)[\Lambda S_0^{-1}{\bar S}_0+{\bar \Lambda} {\bar S}^{-1}_0 S_0]_D$  holds \cite{Cecotti:1987sa}, it can be further rewritten as
\begin{align}
{\cal L}=&-3[S_0{\bar S}_0-\lambda_1 \cala{\bar {\cala}} +(1/2)(\Lambda S_0^{-1}{\bar S}_0+{\bar \Lambda} {\bar S}^{-1}_0 S_0) ]_D \notag \\ &+\xi \left[ \nabla^\alpha (\cala/S_0)  \nabla_\alpha ({\cala}/S_0)  {\bar \nabla}_{\dot \alpha} ({\bar \cala}/{\bar S}_0) {\bar \nabla}^{\dot \alpha} ( {\bar \cala}/{\bar S}_0) \right]_D+3[\Lambda \cala]_F+{\rm h.c.}, 
\end{align}
which will lead to the standard Poincar\'e supergravity with a chiral multiplet that has higher-order derivative coupling. 
Defining new chiral multiplets
\begin{equation}
C\equiv \frac{\sqrt{\lambda_1}\cala}{S_0}, \quad T=\frac{\Lambda}{2 S_0^2}+\frac{1}{2}, 
\end{equation}
we obtain the Lagrangian
\begin{equation}
{\cal L}=-3[S_0{\bar S}_0(T+{\bar T}-C{\bar C})]_D+\frac{\xi}{\lambda_1^2}\left[ \nabla^\alpha C  \nabla_\alpha C {\bar \nabla}_{\dot \alpha} {\bar C}{\bar \nabla}^{\dot \alpha} {\bar C} \right]_D+\frac{6}{\sqrt{\lambda_1}}\left[S_0^3 C\left(T-\frac{1}{2}\right)\right]_F+{\rm h.c.}
\end{equation}
After gauge fixing $S_0=1$, the Lagrangian leads to, 
\begin{align}
{\cal L}=&\int d^2 \Theta 2 {\cal E} \left[\frac{3}{8}\left({\bar {\cal D}} {\bar {\cal D}}-8{\cal R}\right)e^{-K/3}+W \right]+{\rm h.c.} \notag \\
&-\frac{\xi}{\lambda_1^2} \int d^2 \Theta 2 {\cal E} \left[ \frac{1}{8}\left({\bar {\cal D}} {\bar {\cal D}}-8{\cal R}\right) {\cal D}^\alpha C  {\cal D}_\alpha C {\bar {\cal D}}^{\dot \alpha} {\bar C}{\bar {\cal D}}_{\dot \alpha} {\bar C}\right], 
\end{align}
where
\begin{equation}
K\equiv -3\ln[T+{\bar T}-C{\bar C}], \quad W=\frac{6}{\sqrt{\lambda_1}}C\left(T-\frac{1}{2}\right). 
\end{equation}

Expanding the component fields,  integrating out the auxiliary fields in the gravity sector and 
the F-term of the $T$ field,  and performing an appropriate Weyl transformation, we have 
\begin{align}
{\cal L}=\sqrt{-g}&\left[-\frac{R}{2}-K_{i{\bar j}}\partial_\mu z^i \partial^\mu z^{*j}-\frac{12}{\lambda_1}\frac{|C|^2}{T+T^*-|C|^2}\left(1-\frac{3(T+T^*-1)}{T+T^*-|C|^2}\right)\right. \notag \\
&+\frac{6}{\sqrt{\lambda_1}(T+T^*-|C|^2)^2}\left\{\left(|C|^2+T-\frac{1}{2} \right)F_C+{\rm h.c.}\right\}+\left(\frac{3}{(T+T^*-|C|^2)^2}-\frac{32\xi}{\lambda_1^2}\frac{\partial_\mu C \partial^\mu C^*}{T+T^*-|C|^2}\right)|F_C|^2 \notag \\
&\left.+\frac{16 \xi}{\lambda_1^2}\partial_\mu C \partial^\mu C \partial_\nu C^* \partial^\nu C^* +\frac{16\xi}{\lambda_1^2 (T+T^*-|C|^2)^2}|F_C|^4\right],  
\end{align}
where $z^i=C,T$, $K_{i{\bar j}}\equiv \partial^2 K /\partial z^i \partial z^{*j}$ and $F_C$ is the 
F-term of $C$. 
Here we have used the same symbol for the superfield and its scalar component. 
Since $K_{i{\bar j}}$ has the form
\begin{equation}
K_{T{\bar T}}=\frac{3}{(T+T^*-|C|^2)^2}, \quad K_{T{\bar C}}=\frac{-3C}{(T+T^*-|C|^2)^2},\quad K_{C{\bar C}}=\frac{3(T+T^*)}{(T+T^*-|C|^2)^2},
\end{equation}
the system does not have the ghost instability as long as $T+T^*>|C|^2$. 
Note that the equation of motion for $C$ has only up to the second-order derivative, and hence 
there arise no additional degrees of freedom. 

$\partial {\cal L}/\partial F_C=0$ gives the condition that $F_C$ satisfies, 
\begin{equation}
A+B F_C^*+2SF_C F_C^{*2}=0, \label{fcconst}
\end{equation}
where 
\begin{align}
A&=\frac{6}{\sqrt{\lambda_1}(T+T^*-|C|^2)^2}\left(|C|^2+T-\frac{1}{2}\right), \\
B&=\left(\frac{3}{(T+T^*-|C|^2)^2}-\frac{32\xi}{\lambda_1^2}\frac{\partial_\mu C \partial^\mu C^*}{T+T^*-|C|^2}\right), \\
S&=\frac{16\xi}{\lambda_1^2 (T+T^*-|C|^2)^2}. 
\end{align}
Then, $|F_C|^2$ satisfies the equation
\begin{equation}
\alpha=(1+\beta|F_C|^2)^2|F_C|^2,  
\end{equation}
with 
\begin{equation}\alpha=\frac{|A|^2}{B^2}, \quad \beta=\frac{2S}{B}. 
\end{equation}
Here $\alpha$ is always positive, and 
assuming that $\partial_\mu C=0$, the sign of $\beta$ is determined by $\xi$. 

In the case  $\beta>0$, 
Eq.~\eqref{fcconst} has only one real and positive solution, 
\begin{equation}
|F_C|^2=\frac{2}{3 \beta}(\cosh m -1), \label{fterm}
\end{equation}
where
\begin{equation}
m=\frac{1}{3}\cosh^{-1}\left(\frac{27}{2}\alpha \beta+1\right).  
\end{equation}
Note that $1+(27/2)\alpha \beta>1$ is always satisfied in this case. 
On the other hand, in the case $\beta<0$, the situation is relatively complicated. 
If $0<\alpha<-(4/27)\beta^{-1}$, or $-1<1+(27/2)\alpha \beta<1$, Eq.~\eqref{fcconst} has three real and positive solutions, 
\begin{equation}
|F_C|^2=\left\{
\begin{array}{l}
\dfrac{2}{3 \beta}(\cos {\tilde m} -1) \\ \\
\dfrac{2}{3 \beta}\left(\cos \left({\tilde m}+\dfrac{2\pi}{3}\right) -1\right) \\ \\
\dfrac{2}{3 \beta}\left(\cos \left({\tilde m}-\dfrac{2\pi}{3}\right) -1\right) 
\end{array}\right., 
\end{equation}
where
\begin{equation}
{\tilde m}=\frac{1}{3}\cos^{-1}\left(\frac{27}{2}\alpha \beta+1\right).  
\end{equation}
If $\alpha>(4/27)\beta^{-1}$ it has again only one real and positive solution, 
\begin{equation}
|F_C|^2=\frac{1}{3 \beta} (-2+Z^{1/3}+Z^{-1/3}),
\end{equation}
with 
\begin{equation}
Z\equiv\frac{2+27 \alpha \beta+\sqrt{27 \alpha \beta(4+27\alpha \beta)}}{2}. 
\end{equation}

Let us study the resultant Lagrangian. 
The full Lagrangian in which all the auxiliary fields are integrated out is
\begin{align}
{\cal L}=\sqrt{-g}&\left[-\frac{R}{2}-K_{i{\bar j}}\partial_\mu z^i \partial^\mu z^{*j}+\frac{16 \xi}{\lambda_1^2}\partial_\mu C \partial^\mu C \partial_\nu C^* \partial^\nu C^* \right. \notag \\
&\left.-\frac{12}{\lambda_1}\frac{|C|^2}{T+T^*-|C|^2}\left(1-\frac{3(T+T^*-1)}{T+T^*-|C|^2}\right)-B|F_C|^2-3S|F_C|^4\right], \label{fulllag}
\end{align}
independent of the value of $\beta$ with F terms given above. 
Taking $C=0$\footnote{For small $\xi$, the Lagrangian \eqref{fulllag} 
reveals a tachyonic instability for $C$ \cite{Farakos:2013cqa}. However, by introducing 
the $[({\cal R}{\bar {\cal R}})^2/(S_0{\bar S}_0)]_D \rightarrow [(C{\bar C})^2]_D$ term, $C$ can acquire a positive mass squared
and the tachyonic instability problem can be solved \cite{Kallosh:2013lkr,Farakos:2013cqa}. 
Here we assume implicitly such an extra term. 
Note that such a term does not change the Lagrangian for $T$ in the $C=0$ direction. We discuss it
in more detail in Appendix~\ref{app2}. }, 
the Lagrangian is now of the form
\begin{align}
{\cal L}=\sqrt{-g}&\left[-\frac{R}{2}-\frac{3}{(T+T^*)^2}\partial_\mu T \partial^\mu T^*-B|F_C|^2-3S|F_C|^4\right]. 
\end{align}
Let us define
\begin{equation}
T=\frac{1}{2}e^{\sqrt{2/3}\phi }+ ib, 
\end{equation}
to canonicalize the real part of the $T$ field. 
Noting that we now have
\begin{align}
B&=\frac{3}{(T+T^*)^2}=3 e^{-2 \sqrt{2/3}\phi}, &S&=\frac{16 \xi}{\lambda_1^2(T+T^*)^2}=\frac{16 \xi}{\lambda_1^2} e^{-2 \sqrt{2/3}\phi}, \notag \\
\alpha&=\frac{4}{\lambda_1}\left|T-\frac{1}{2}\right|^2=\frac{(e^{\sqrt{2/3}\phi}-1)^2+4b^2}{\lambda_1}, &\beta&=\frac{32\xi}{3\lambda_1^2}, 
\end{align}
the Lagrangian becomes
\begin{align}
{\cal L}=\sqrt{-g}&\left[-\frac{R}{2}-\frac{1}{2}\partial_\mu \phi  \partial^\mu \phi-3e^{-2\sqrt{2/3}\phi}\partial_\mu b \partial^\mu b -V\right]
\end{align}
where 
\begin{equation}
V(\phi)=\frac{3\lambda_1^2}{16 \xi} e^{-2\sqrt{2/3}\phi} X(X-1) \label{potential}
\end{equation}
with 
\begin{equation}
|F_C|^2=\frac{2}{3 \beta}(X-1). 
\end{equation}
The expression of $X$ is different depending on the values of $\xi$ and $\phi$ as
\begin{equation}
X= \left\{
\begin{array}{ll}
\cosh m & \text{for} \quad \xi>0, \\
\left(
\begin{array}{l}
\cos {\tilde m} \\
\cos \left({\tilde m}+2\pi/3\right) \\
\cos \left({\tilde m}-2\pi/3\right) 
\end{array} \right. & \text{for} \quad \xi<0 \quad \text{and} \quad  \phi<\sqrt{\dfrac{3}{2}}\log \left[ 1+\dfrac{1}{6 \sqrt{2|s|}}\right] \equiv \phi_c \\
(Z^{1/3}+Z^{-1/3})/2 & \text{for} \quad \xi<0 \quad \text{and} \quad  \phi >\phi_c
\end{array} \right. , \label{XX}
\end{equation}
where 
\begin{equation}
s\equiv \frac{\xi}{\lambda_1^3}. 
\end{equation}
Here, the $X=\cos({\tilde m}+2\pi/3)$ branch for $\xi<0$ smoothly connects to the solution for $\phi>\phi_c$. 
$m$ and ${\tilde m}$ are expressed by $\phi$ and $b$ as
\begin{align}
m&=\frac{1}{3}\cosh^{-1}\left[144s((e^{\sqrt{2/3}\phi}-1)^2+4b^2)+1\right], \label{m1} \\
{\tilde m}&=\frac{1}{3}\cos^{-1}\left[144 s((e^{\sqrt{2/3}\phi}-1)^2+4b^2)+1\right],  \label{m2}  
\end{align}
and the condition $\alpha <-(4/27)\beta^{-1}$ yields $\phi<\phi_c$. 

One may wonder which branch to take for $s<0$. 
We find that the branch $X= \cos {\tilde m}$ has the potential minimum $V=0$ at $\phi=0$ for $b=0$ and 
approaches the pure Starobinsky model for the $s \rightarrow 0$ limit, 
whereas other branches as well as the solution $\phi>\phi_c$ 
have no potential minimum, and 
the potential takes  a negative value  at $\phi \rightarrow -\infty$. 
Therefore, we take the branch $X= \cos {\tilde m}$  as the supresymmetrized Starobinsky model 
with an $R^4$ correction for $s<0$ and $\phi<\phi_c$. 
Since other branches do not have well-defined vacua, hereafter we do not consider them. 
In Appendix~\ref{app}, we show that the resultant potential is the equivalent to the 
Starobinsky model with a $R^4$ correction in the nonsupersymmetric case, which strongly suggests that the model is 
its supersymmetrized one. 

\begin{figure}[t]
\center
\includegraphics[width = 0.8\textwidth]{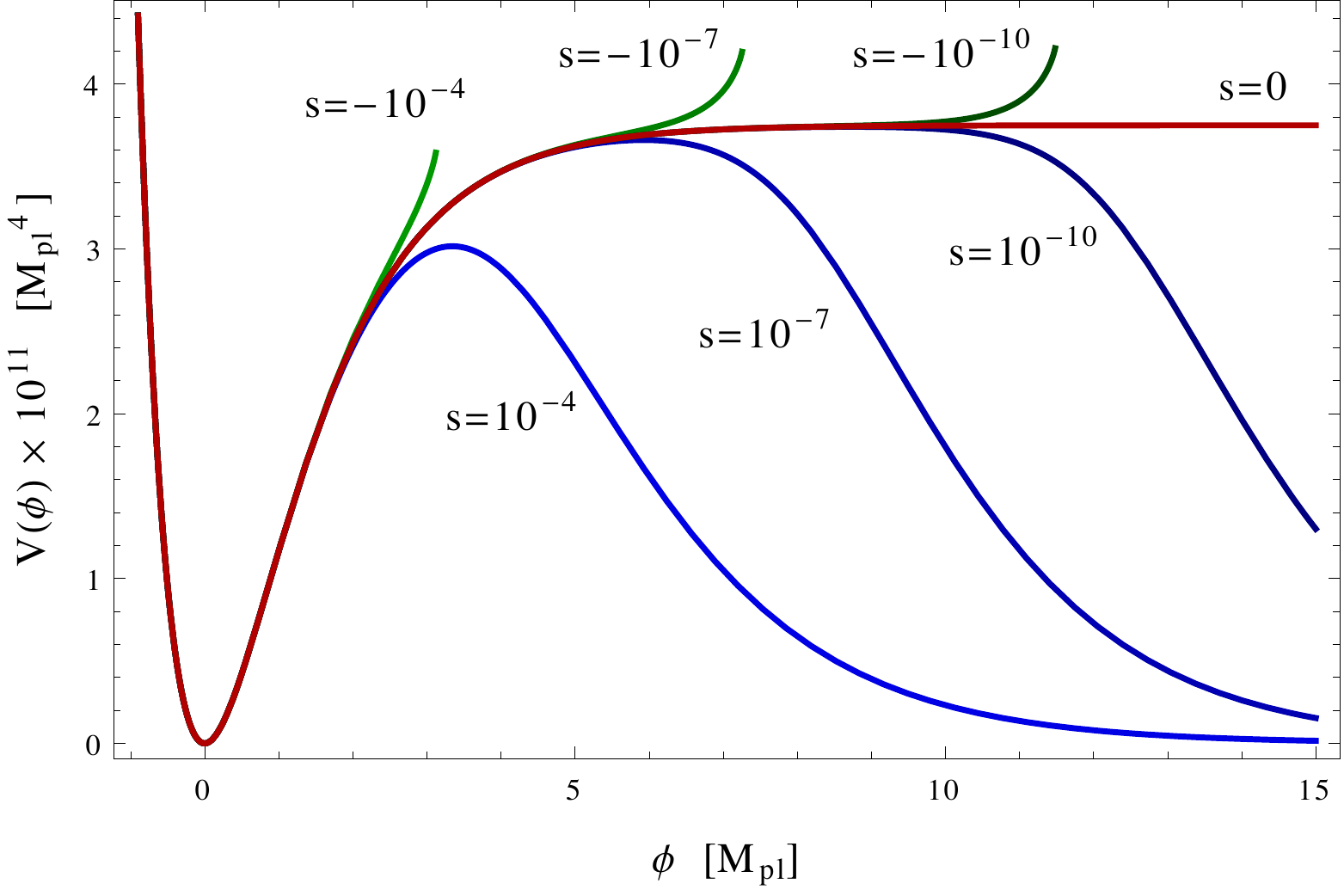}
\caption{The potential with higher-order correction for the Starobinsky model with various amplitudes of the corrections is shown. 
Here we take $\lambda_1=8 \times 10^{10}(M_{\rm pl}^{-2})$.}
\label{fig:1}
\end{figure}

Figure \ref{fig:1} shows the parameter dependence of the potential shape for $b=0$. 
For $s>0$, the potential has a maximum and approaches to $V=0$ at $\phi=0$ and $\phi \rightarrow \infty$. 
On the other hand, for $s<0$, it is a continuously increasing function with respect 
to $\phi$ and undefined for $\phi>\phi_c$. 
In both cases, the flatness of the potential appears to be violated for $|s|>10^{-7}$, 
making it difficult for inflation to take place. 
We will see how inflation can take place and how the correction is constrained observationally 
in the next section. 

Here we comment on the $b$ field. 
Since the imaginary part $b$ receives a positive mass squared for $\phi>0$ larger than $H^2$ along the 
inflationary trajectory, 
we can safely take $b=0$ and we have only to focus on the dynamics of $\phi$ field. 
In Appendix~\ref{app2}, we examine it in detail.

\section{Observational constraints \label{sec3}}

Now, let us consider inflation driven by the $\phi$ field and study the observational constraints on the model parameters. 
Inflation takes place when  the slow-roll conditions, 
\begin{equation}
\epsilon \equiv \frac{1}{2}\left(\frac{\partial V/\partial \phi}{V}\right)^2 \ll 1, \quad |\eta| \equiv \frac{1}{V}\left|\frac{\partial^2 V}{\partial \phi^2}\right| \ll1,  
\end{equation}
are satisfied. 
We can easily find that, even for relatively large $|s| \gg 1$, there are field spaces in which  
slow-roll conditions $\eta \ll 1, \epsilon \ll 1 $ are simultaneously satisfied. 
Therefore, slow-roll inflation can take place naturally in a sense apart from the observational consequences
and initial condition problem.
We will turn to the latter in the next section. 
During inflation, the system obeys the slow-roll equations, 
\begin{equation}
3H^2=V(\phi), \quad \frac{d \phi}{dN}=\frac{\partial V/\partial \phi}{V}, 
\end{equation}
where 
$N$ is defined as $dN=-Hdt$, and inflation ends when 
$\epsilon$ reaches unity 
at $\phi=\phi_{\rm f}$. 

Cosmological perturbations are generated during inflation. They are quantified in terms of 
the amplitude of the scalar fluctuations $A_s$, the scalar spectral index $n_s$ and the tensor-to-scalar ratio $r$, 
\begin{align}
A_s&= \frac{H^2}{8\pi \epsilon}, \label{asas} \\
n_s&=1-6 \epsilon+2 \eta, \\
r&=16 \epsilon,  
\end{align}
which are evaluated at the $\phi$ field value when the relevant scale leaves the horizon during inflation. 
If there are no additional sources of cosmological perturbations, 
they are directly compared to  the Planck and other cosmological observations. 
We adopt the $\phi$ field value at the number of $e$-folds at $N_*\simeq  55$ before the end of inflation when the pivot scale 
$k_*=0.05 \rm{Mpc}^{-1}$ leaves the horizon\footnote{See Ref.~\cite{Audren:2014cea} for the discussion 
of the pivot scale in light of the BICEP2 result.}. 

The model parameters, $\lambda_1$ and $\xi$, are constrained by the observations \cite{Ade:2013uln, Audren:2014cea}, 
\begin{equation}
A_s^{\rm obs} =(2.18 \pm 0.05)\times 10^{-9}, \quad n_s^{\rm obs} =0.963 \pm 0.007, \quad r<0.135. \label{planckres}
\end{equation} 
For $s=0$, the scalaron potential\footnote{The dynamics of scalaron
oscillation with this potential is investigated in Ref.~\cite{Takeda:2014qma}
.} is given by 
\begin{equation}
V(\phi)=\frac{3}{\lambda_1}(1-\exp[-\sqrt{3/2}\phi])^2, 
\end{equation}
which yields $3 H^2 \simeq 3/\lambda_1$ during inflation, 
and solving the slow-roll equations analytically, we obtain 
\begin{equation}
\epsilon \simeq \frac{3}{4N_*^2}. 
\end{equation}
Then, comparing Eqs.~\eqref{asas} and \eqref{planckres}, we find 
\begin{equation}
\lambda_1 =\frac{N_*^2}{6 \pi A_s^{\rm obs}} \simeq 7 \times 10^{10}.
\end{equation}

\begin{figure}[t]
\centering{
\includegraphics[width = 0.45\textwidth]{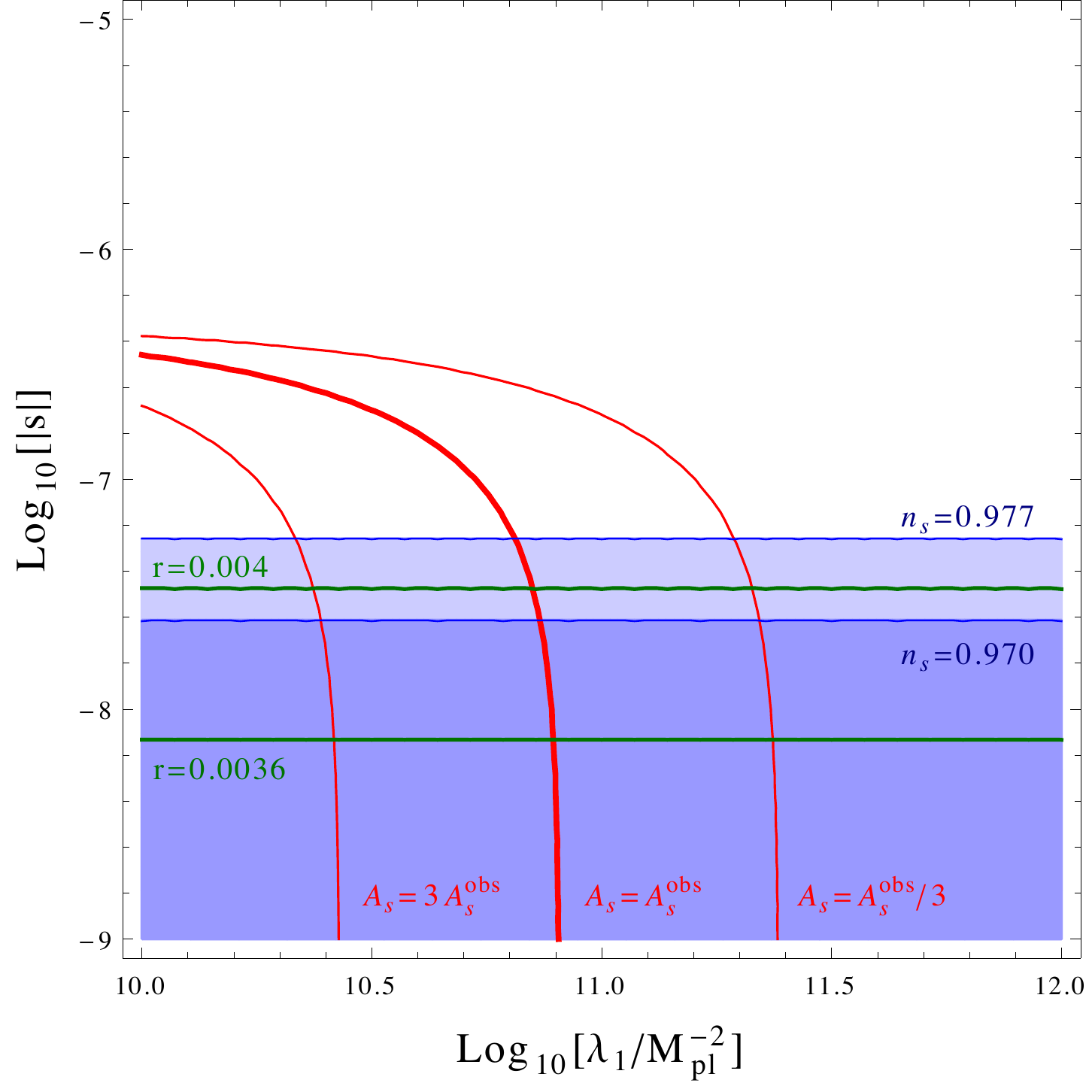}
\includegraphics[width = 0.45\textwidth]{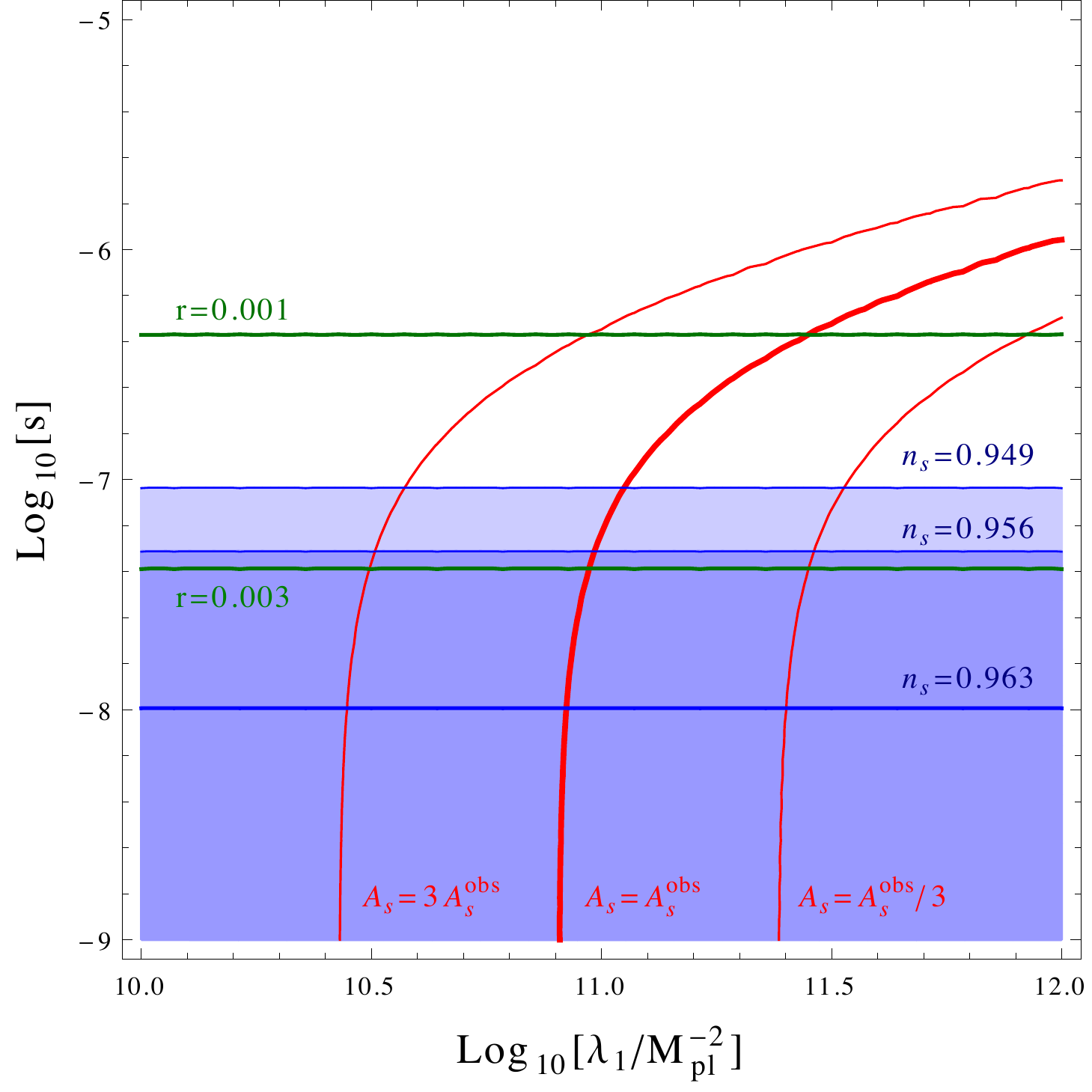}
}
\caption{The cosmological perturbations generated from the supersymmetric Starobinsky model 
with higher-order corrections with respect to the model parameters are shown;  
({\it left}) in the case with $s<0$ and ({\it right}) in the case with $s>0$. 
Curved (red) lines represent the contour lines of the amplitude of the scalar fluctuation $A_s$. 
The thick (light) shaded (blue) regions are the regions in which the values of the scalar spectral index are within the 
Planck $1 \sigma (2\sigma)$ constraints. Horizontal (green) lines are the contour lines for the typical values of 
the tensor-to-scalar ratio $r$. }
\label{fig:2}
\end{figure}

Since the slow-roll dynamics of $\phi$ cannot be solved analytically when  the the higher-order 
correction exists, we have performed numerical calculation of the inflationary dynamics and evaluated the 
primordial perturbations with various values of $\lambda_1$ and $s$.
Parameter regions that are favored by Planck are shown in Fig. \ref{fig:2} in the cases $s<0$ and $s>0$. 
In both cases, for $|s|>10^{-7}$, the higher-order corrections are no longer negligible 
for the inflaton dynamics, and hence the predictions start to deviate from the 
pure Starobinsky model's; for $A_s=A_s^{\rm obs}$, $\lambda_1\simeq 7 \times 10^{10} $ is required, 
and for these parameter values, the scalar spectral index and the tensor-to-scalar ratios are 
predicted as $n_s \simeq 0.963$ and $r\simeq 3\times 10^{-3}$. 
As a result, the value of $s$ is constrained as
\begin{equation}
-5.5 \times 10^{-8} < s<9.1\times 10^{-8},
\end{equation}
by the Planck observation at the $2 \sigma$ confidence level. 
Therefore we conclude that the amplitude of the parameter $s$ must be smaller than at least 
$10^{-7}$ to explain the current Universe in the context of the supergravity
Starobinsky model. 
This means that some symmetries or mechanisms to reduce the higher-order corrections 
to the Starobinsky model  up to this level are necessary to derive it from the physics in the higher energy scales. 

Here, we explain the behaviors of the parameter dependence of the observables. 
In the case with $s<0$, larger $|s|$ leads to  larger values of $\epsilon$ during inflation. 
Therefore, larger potential energy or smaller $\lambda_1$ is required to generate the correct amplitude 
of the scalar perturbations $A_s$, which leads to relatively large values of the tensor-to-scalar ratio $r$. 
At the same time, the slow-roll parameter $\eta$ becomes a larger or even positive value, which leads to a 
larger value of the scalar spectral index $n_s$. 
On the other hand, in the case  with $s>0$, larger $s$ leads to  smaller values of $\epsilon$ during inflation. 
This leads to  larger values of $\lambda_1$ to generate the correct amplitude of $A_s$, which 
also means  smaller values of $r$. Simultaneously, for larger $s$, $\eta$ becomes larger, 
which leads to a smaller value of $n_s$. 
Note that the slow-roll parameters are independent of $\lambda_1$, and hence the observables 
$n_s$ and $r$ are $\lambda_1$ independent. 

\section{Initial condition for inflation \label{sec4}}

Now, let us consider the initial condition problem, which is
strikingly different depending on the sign of $s$. 
First, for $s<0$, as we have seen above, the field range that can evolve
into the proper vacuum after inflation is limited to $\phi < \phi_c$.
For a sufficient amount of inflation, slow-roll inflation should start
at $\phi \gtrsim 6$. Hence, severe fine-tuning of the initial condition
at, say, the Planckian epoch is necessary for both $\phi$ and $\dot\phi$.
If $\dot\phi$ has a Planckian value $\dot\phi \sim 1$
initially, the scalar field amplitude varies $\Delta\phi \sim 10$ before
 the slow-roll inflation phase sets in. 
Therefore, $\phi_c$ must be larger than $15-16$ for this initial velocity, 
which turns to the constraint on $s$ as $|s|<3.2 \times 10^{-13}$. 
For the larger amplitude of $s$, 
the initial velocity must be suppressed accordingly. 

On the other hand, for $s>0$, there is no restriction in the field range
of $\phi$, and the potential has a local maximum at $\phi \equiv \phi_t
\simeq -0.93\log_{10}(3.5s)$.  
Hence, if the universe starts with a chaotic initial condition, some
domain falls into  $\phi=0$, and others run away to infinity. 
Note that since the $C$ and $b$ fields are stabilized at the origin for 
any field values of $\phi$ 
we can take $C=b=0$ in all the domains.
Between these domains with different fates exists a region trapped to the potential maximum at $\phi=\phi_t$, 
namely, a domain wall.  
As Vilenkin and Linde have pointed out \cite{Linde:1994hy}, inflation can naturally take place 
inside the domain wall, where the large energy density distributes relatively homogeneously, 
if its thickness is larger than the local Hubble radius. 
This is so-called the ``topological inflation.'' 
Thus, in the present case, it may be possible for the topological 
inflation to take place. 

\begin{figure}[t]
\centering{
\includegraphics[width = 0.45\textwidth]{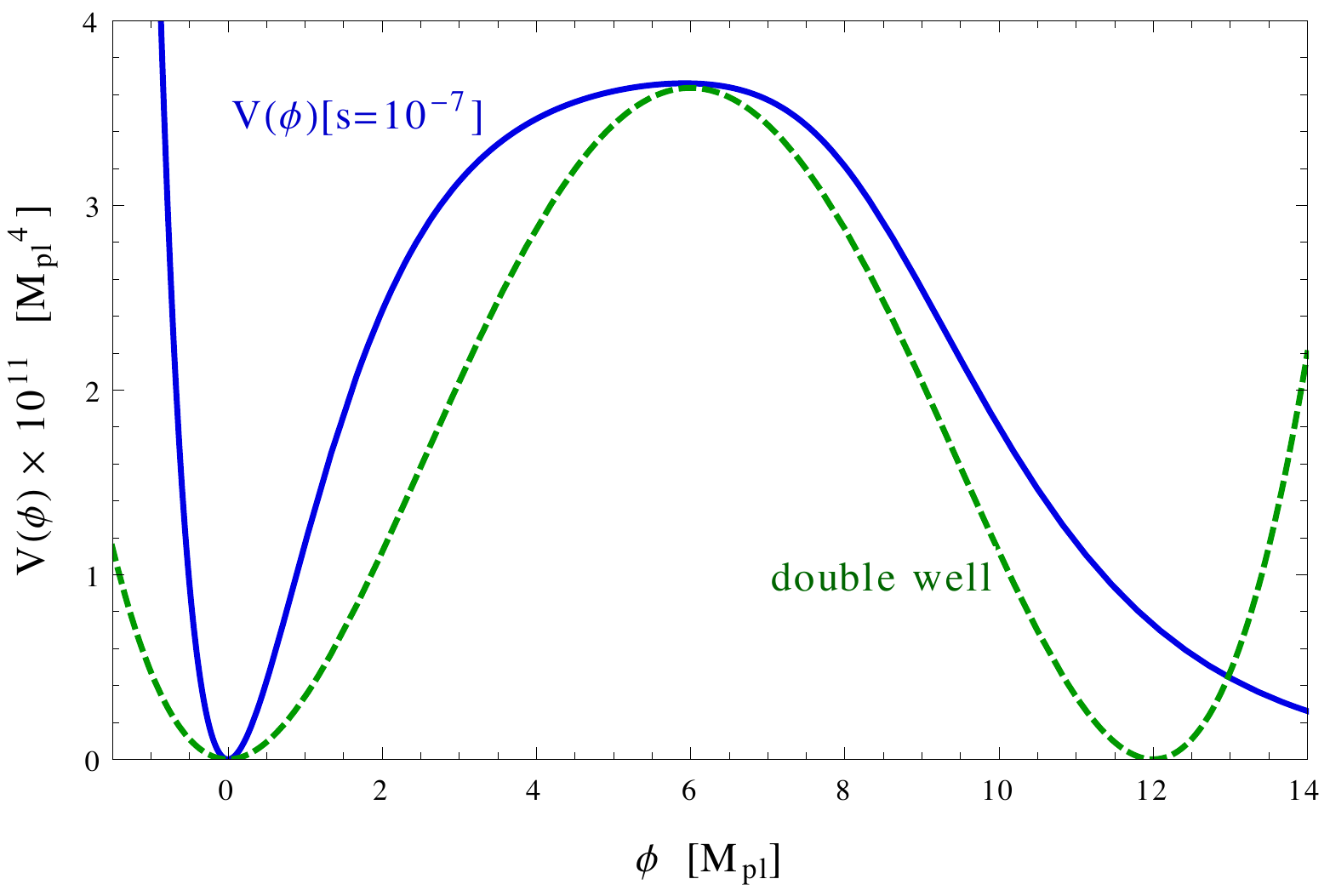}
\includegraphics[width = 0.45\textwidth]{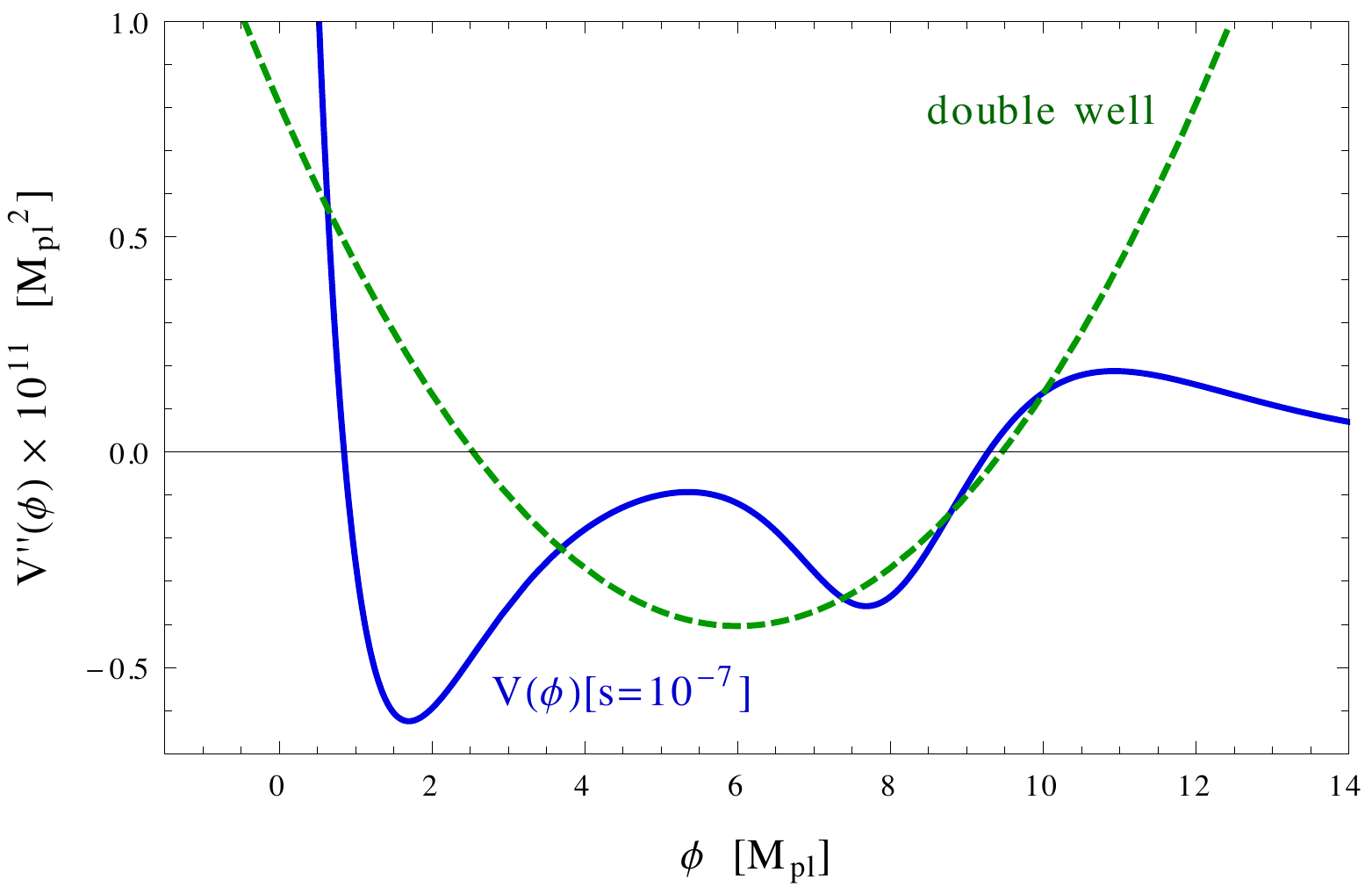}
}
\caption{The potential ({\it left}) and its second derivative with respect to $\phi$ ({\it right}) of both the Starobinsky model 
with higher order corrections $s=10^{-7}$ and the double well potential are shown. 
The double potential has the maximum at $\phi=6$ and the amplitude is the same to the Starobinsky model. 
Around the potential maximum, the Starobinsky model is flatter than the double-well potential. }
\label{fig:5}
\end{figure}

Let us study this possibility in detail. 
The condition for the realization of topological inflation is numerically studied in Ref.~\cite{Sakai:1995nh}
in the case with a potential $V=\kappa(\phi^2-v^2)^2$, and it was concluded that 
a domain wall triggers inflation if $v$ is larger than the critical value $v_c \equiv 1.7$ regardless of 
the value of $\kappa$. Since the potential we are studying is different from the double-well type, 
the conclusion in Ref.~\cite{Sakai:1995nh} cannot be applied directly. 
However, it is plausible that inflation can take place from the domain wall in the following reasons. 
The thickness of the domain wall can be evaluated as $\delta = |V''(\phi_t)|^{-1/2}$. 
Since numerically we find that 
\begin{equation}
V''(\phi_{\rm t})\simeq -2.1 \times 10 \frac{s^{1/3}}{\lambda_1}
\end{equation} 
and the Hubble parameter is evaluated as $H= (V(\phi_{\rm t})/3)^{1/2} \simeq 1/\lambda_1^{1/2}$, 
we have the relation 
\begin{equation}
H \delta \simeq 0.22 s^{-1/6}. 
\end{equation}
Since the condition for the critical ratio between the wall thickness to the Hubble length given in Ref.~\cite{Sakai:1995nh} is 
$H \delta =0.48$, the one in our case for $s<10^{-7}$ is much larger than the critical value. 
Figure~\ref{fig:5} shows the shape of potential and its second derivative both in the case with our potential 
with $s=10^{-7}$ and the double-well potential with $v\simeq 6$, both of which have the same potential maximum. 
We can see that our potential is flatter than the double-well potential that can trigger the topological inflation. 
Therefore, topological inflation will naturally take place in our potential
satisfying the observational constraint $s < 9.1 \times 10^{-8}$. 

\section{summary \label{sec5}}

Starobinsky's $R^2$ inflation is one of the most attractive inflation models in light of the 
Planck result. However, 
the mechanism to induce the correct $R^2$ term that explains the observational result is not known. 
Therefore, it would be a good direction to embed it in a supersymmetric theory 
because it is one of the most promising physics beyond the Standard Model 
and would be the key to the quantum theory of gravity. 
On the other hand, once we consider the supersymmetric theory of the $R^2$ model, 
higher-order terms cannot be forbidden by symmetry. 

In this paper, we have studied the Starobinsky model in the old-minimal supergravity 
with an $R^4$ correction that is free from ghost degrees of freedom. 
After confirming that fields other than the scalaron field are stabilized appropriately, we focused on the dynamics of the scalaron sector.
Since the $R^4$ correction easily violates the flatness of the inflaton potential 
in the scalaron picture, it should be strongly constrained. 
We find that the constraint on the $R^4$ term is not so strong just for 
the accelerating expansion of the Universe, but in order to generate the 
spectral index of primordial scalar perturbation that is consistent with 
Planck result, it is strongly constrained. 
It is found that in terms of dimensionless coupling constant $s \equiv \xi/\lambda_1^3$  
it is constrained as 
\begin{equation}
-5.5 \times 10^{-8} <s < 9.1 \times 10^{-8}. 
\end{equation}

On the initial condition, we also find the difficulties in the realization of inflation
when there is an $R^4$ correction. 
From the chaotic initial condition in which the Universe starts from the Planck scale, 
the $R^4$ term must be very severely constrained for $s<0$, 
where the scalaron potential jumps up at the field value larger than the value 
at which the $R^4$ correction becomes dominant. 
On the other hand, in the case of $s>0$, the shape of the scalaron potential is 
hilltop type, and domain walls are generated somewhere in the Universe regardless of the initial condition. 
We find that the domain wall is thick enough for the topological inflation 
for $s<10^{-7}$, and hence we do not suffer from the initial condition problem 
in this case. 
In summary, for the reasonable initial conditions, the $R^4$ correction is 
constrained as 
\begin{equation}
-3.2 \times 10^{-13}\ll s < 9.1 \times 10^{-8}
\end{equation}
for the realization of inflation that leads to the present Universe. 
This would be an important constraint for the embedding or inducing 
the Starobinsky model of inflation from the high-energy theory.

\begin{acknowledgments}
K.K. is grateful to K.~Ohashi, R.~Rattazzi, and  A.~Westphal for useful comments. 
The work of K.K. is supported by a JSPS postdoctoral fellowship for
research abroad.  The work of J.Y. is supported by JSPS 
Grant-in-Aid for Scientific Research, Grant No.~23340058.

\end{acknowledgments}
\appendix
\section{Higher-order corrections to the nonsupersymmetric Starobinsky model \label{app}}

Here we examine the higher-order correction to the nonsupersymmetric Starobinsky model  in the scalaron picture.  
We will find that it is strongly suggested that the model we studied is truly its supersymmetrized one. 
Let us consider the following action: 
\begin{equation}
S=-\int d^4 x \sqrt{-g} \frac{R}{2}\left(1-\frac{\bar \lambda}{2}R-\frac{\bar \xi}{4}R^3\right). 
\end{equation}
The equivalent action is
\begin{align}
S&=-\int d^4 x \sqrt{-g}\left\{\frac{\varphi}{2}\left(1-\frac{\bar \lambda}{2}\varphi - \frac{\bar \xi}{4}\varphi^3\right)+\frac{1}{2}(1-{\bar \lambda} \varphi -{\bar \xi} \varphi^3)(R-\varphi)\right\} \notag \\
&=-\int d^4 x \sqrt{-g}\left\{ \frac{1}{2}(1-{\bar \lambda} \varphi -{\bar \xi} \varphi^3)R + \frac{\bar \lambda}{4}\varphi^2+\frac{3\xi}{8} \varphi^4\right\}. 
\end{align}
Performing the conformal transformation $g_{\mu\nu}\rightarrow {\bar g}=\Omega^2 g_{\mu\nu}$ with 
$\Omega^2=1-{\bar \lambda} \varphi-{\bar \xi} \varphi^3$, the action becomes
\begin{equation}
S=-\int d^4 x \sqrt{-{\bar g}} \left\{\frac{\bar R}{2}+\frac{3}{4}\left(\frac{{\bar \lambda}+3 {\bar \xi} \varphi^2}{1-{\bar \lambda}\varphi-{\bar \xi}\varphi^3}\right)^2 \partial_\mu \varphi \partial^\mu \varphi + \frac{1}{(1-{\bar \lambda}\varphi-{\bar \xi}\varphi^3)^2}\left(\frac{\bar \lambda}{4}\varphi^2+\frac{3{\bar \xi}}{8} \varphi^4\right)\right\}. 
\end{equation}
Defining 
\begin{equation}
\chi\equiv \sqrt{\frac{3}{2}}\log\left[1-{\bar \lambda}\varphi-{\bar \xi}\varphi^3\right], \label{nonsusycan}
\end{equation}
we have the action for the canonically normalized field $\chi$, 
\begin{equation}
S=-\int d^4 x \sqrt{-{\bar g}} \left\{\frac{\bar R}{2}+\frac{1}{2} \partial_\mu \chi \partial^\mu \chi + e^{-2\sqrt{2/3}\chi}\left(\frac{\bar \lambda}{4}\varphi^2[\chi]+\frac{3{\bar \xi}}{8} \varphi^4[\chi]\right)\right\}. 
\end{equation}
Eq.\eqref{nonsusycan} can be solved as
\begin{equation}
\varphi^2[\chi]=\left\{
\begin{array}{ll}
\dfrac{2{\bar \lambda}}{3{\bar \xi}}(\cosh {\bar m}_1[\chi]-1) & \text{for} \quad \xi>0, \\
\dfrac{2{\bar \lambda}}{3{\bar \xi}}(\cos {\bar m}_2[\chi]-1) & \text{for} \quad \xi<0,
\end{array}\right.
\end{equation}
with
\begin{align}
{\bar m}_1[\chi]&=\frac{1}{3}\cosh^{-1} \left(\frac{27{\bar \xi}}{2{\bar \lambda}^3}(e^{\sqrt{2/3}\chi}-1)^2+1\right) \quad \text{for} \quad \xi>0, \\
{\bar m}_2[\chi]&=\frac{1}{3}\cos^{-1} \left(\frac{27{\bar \xi}}{2{\bar \lambda}^3}(e^{\sqrt{2/3}\chi}-1)^2+1\right) \quad \text{for} \quad \xi<0. 
\end{align}
Again, for $\xi<0$, there are three solutions for $\varphi[\chi]$, and here we take the solution that approaches the 
Starobinsky model in the $\xi\rightarrow 0$ limit.  
Therefore, the potential for $\chi$ is expressed as
\begin{equation}
V(\chi)=\frac{{\bar \lambda}^2}{6{\bar \xi}}e^{-2\sqrt{2/3}\chi} \times \left\{
\begin{array}{ll}
\left(\cosh {\bar m}_1[\chi]-1\right) \cosh {\bar m}_1[\chi] & \text{for} \quad \xi>0, \\
\left(\cos {\bar m}_2[\chi]-1\right) \cos {\bar m}_2[\chi] & \text{for} \quad \xi<0.
\end{array}\right.
\end{equation}
Comparing them with Eqs.~\eqref{potential}, \eqref{XX}, \eqref{m1}, and \eqref{m2}, 
we find that they are equivalent with the relation
\begin{equation}
\lambda_1=12{\bar \lambda}, \quad \xi=162{\bar \xi}. 
\end{equation}

\section{Masses of $C$ and $b$ fields \label{app2}}

In this appendix, we examine the effective mass of $C$ and $b$ fields and 
show that they can be safely stabilized during inflation. 

\subsection{$C$ field} 
In the Starobinsky limit $\xi \rightarrow 0$, the Lagrangian for $C$ becomes
\begin{align}
{\cal L}\ni&-\frac{3(T+T^*)}{(T+T^*-|C|^2)^2}|\partial_\mu C|^2 -\frac{12}{\lambda_1}\frac{|C|^2}{T+T^*-|C|^2}\left(1-\frac{3(T+T^*-1)}{T+T^*-|C|^2}\right) \notag \\
&-\frac{12}{\lambda_1(T+T^*-|C|^2)^2} \left||C|^2+T-\frac{1}{2}\right|^2.  
\end{align}
The mass term for the $C$ field can be read off as
\begin{equation}
V(T,C) \ni \frac{12}{\lambda_1}\frac{1-2(T(T-1)+{\rm h.c.})}{2(T+T^*)^3} |C|^2. 
\end{equation}
Therefore, the $C$ field becomes tachyonic for $T>(\sqrt{2}+1)/2$, neglecting the imaginary part of $T$, 
which may violate the successful inflation. 
This problem is resolved by introducing higher-order term like 
$[\zeta ({\cal R}{\bar {\cal R}})^2/(S_0{\bar S}_0)]_D \rightarrow [\zeta(C{\bar C})^2]_D$ 
with $\zeta$ being a numerical constant.  
This term gives an additional mass term, 
\begin{equation}
\Delta V=-\frac{12\zeta}{\lambda_1} \frac{|2T-1|^2}{(T+T^*)^2}|C|^2. 
\end{equation}
Then, the mass squared of $C$ becomes always positive for $\zeta\ll -0.1$. 
For the fixed $T$, around $C=0$ one can canonically normalize $C$ by multiplying $\sqrt{(T+T^*)/3}$. 
Noting that during inflation $H^2 \simeq V(T)/3$, the ratio between the effective mass squared of $C$ and 
the Hubble parameter becomes
\begin{equation}
\frac{m_{C,{\rm eff}}^2}{H^2}\simeq \frac{1-2(T(T-1)+{\rm h.c.})}{|T-1/2|^2}-4\zeta(T+T^*). 
\end{equation}
Therefore, since during inflation $T+T^*=\exp[\sqrt{2/3}\phi] \simeq 50$, 
the $C$ field is safely stabilized for $\zeta\ll -0.01$. 

\begin{figure}[t]
\center
\includegraphics[width = 0.7\textwidth]{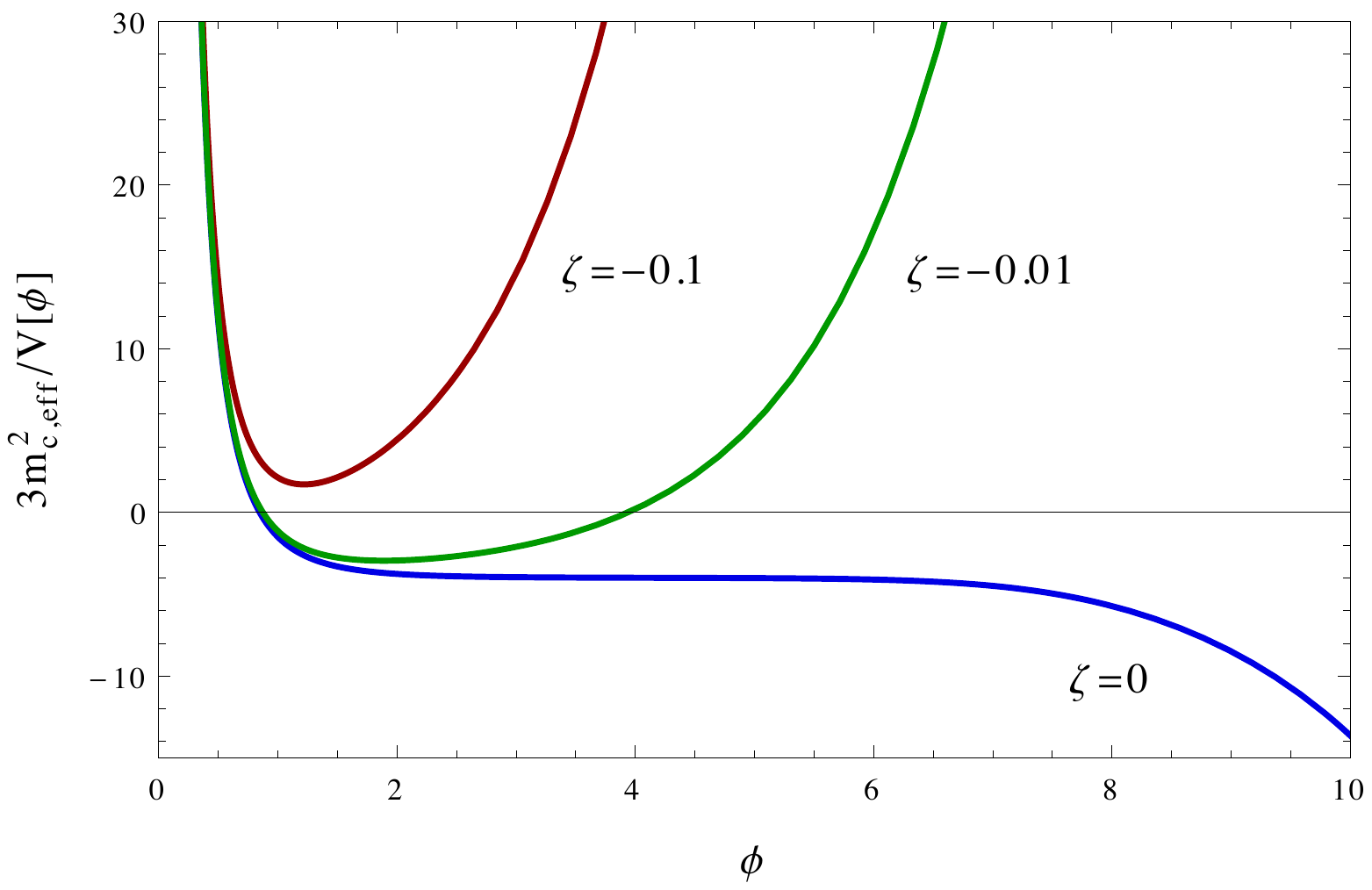}
\caption{The ratio between the effective mass squared of the $C$ field and $V/3 =H^2$ is shown as a function 
of inflaton $\phi$ with $\zeta=0, -0.01$, and -0.1 and $s=10^{-7}$. The $C$ field is safely stabilized 
for $\zeta<-0.1$. }
\label{fig:app1}
\end{figure}

For the nonzero $R^4$ corrections, the situation does not change. 
Figure.~\ref{fig:app1} shows $m_{C,{\rm eff}}^2/(V(\phi)/3)$ with $\zeta=0,-0.01$, and -0.1 
and $\xi=10^{-7}$ as a function of inflaton $\phi$. We can easily see that the $C$ field is safely stabilized 
for $\zeta<-0.1$. The conclusion is the same for $s<0$. 

\subsection{$b$ field}
In the Starobinsky limit $\xi \rightarrow 0$, the Lagrangian for the $b$ field becomes
\begin{equation}
{\cal L} \ni -3 e^{-2 \sqrt{2/3} \phi} \partial_\mu b \partial^\mu b -\frac{12}{\lambda_1}e^{-2 \sqrt{2/3} \phi}  b^2,
\end{equation}
neglecting the $C$ field. 
For the fixed value of $\phi$, the $b$ field is canonically normalized by multiplying $e^{ \sqrt{2/3} \phi}/\sqrt{6}$, and  
the effective mass is read as 
\begin{equation}
m_{b,{\rm eff}}^2=\frac{4}{\lambda_1}. 
\end{equation}
The Hubble parameter during inflation is $H^2=V/3\simeq 1/\lambda_1$, and hence 
the $b$ field is safely stabilized.

\begin{figure}[t]
\center
\includegraphics[width = 0.7\textwidth]{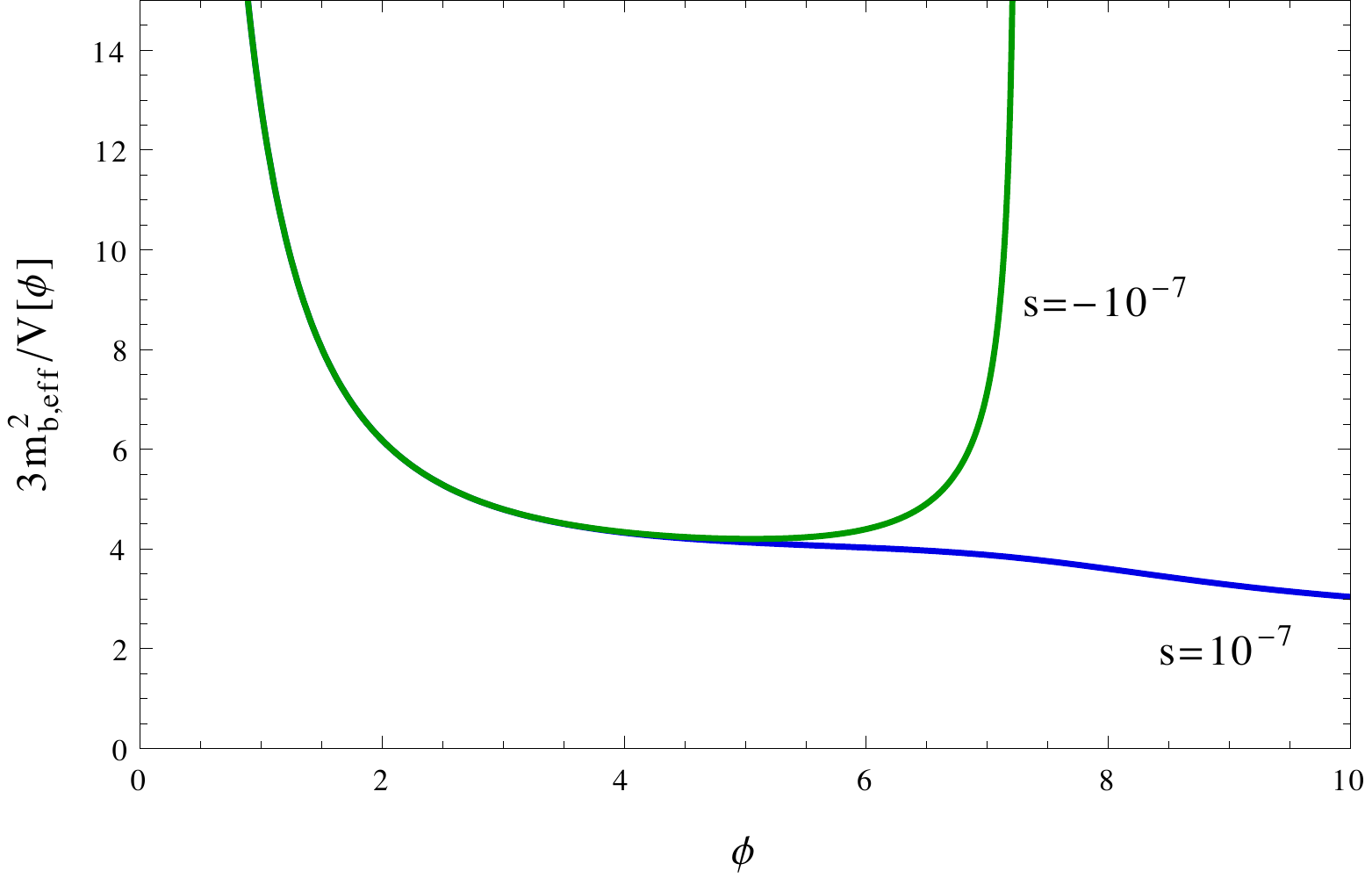}
\caption{The ratio between the effective mass squared of the $b$ field and $V/3 =H^2$ is shown as a function 
of inflaton $\phi$ with $s=10^{-7}$ and $-10^{-7}$.  The $b$ field is safely stabilized. }
\label{fig:app2}
\end{figure}

For the nonzero $R^4$ corrections, again, the situation does not change. 
Figure.~\ref{fig:app2} shows $m_{b,{\rm eff}}^2/(V(\phi)/3)$ with $s=10^{-7}$ and $-10^{-7}$ 
as a function of inflaton $\phi$. We can easily see that the $b$ field is safely stabilized during inflation.

\end{document}